\documentclass[aps,prl,showpacs,twocolumn,superscriptaddress,floatfix]{revtex4-2}
\usepackage[utf8]{inputenc}
\newcommand{\figurescale}{1}
\usepackage{verbatim}
\usepackage{amsmath}
\usepackage{mathtools}
\usepackage{upgreek}
\usepackage{gensymb}
\usepackage[pdftex]{graphicx}
\usepackage[separate-uncertainty=true]{siunitx}
\usepackage[colorlinks=true,urlcolor=blue,linkcolor=blue,citecolor=blue]{hyperref}
\usepackage[usenames,dvipsnames]{xcolor}
\usepackage[T1]{fontenc}
\usepackage{placeins}
\usepackage{nicefrac}
\usepackage{xcolor}
\usepackage[normalem]{ulem}

\usepackage{lineno}

\begin{document}
\title{The effect of surface oxidation and crystal thickness on magnetic properties and magnetic domain structures of Cr$_2$Ge$_2$Te$_6$}

\author{Joachim Dahl Thomsen}
\affiliation{Division of Physical Sciences, College of Letters and Science, University of California, Los Angeles, California 90095, USA}
\affiliation{Department of Materials Science and Engineering, Massachusetts Institute of Technology, Cambridge, Massachusetts 02139, USA}

\author{Myung-Geun Han}\email{mghan@bnl.gov}
\affiliation{Condensed Matter Physics and Materials Science Division, Brookhaven National Laboratory, Upton, NY 11973, USA}

\author{Aubrey Penn}
\affiliation{MIT.nano, Massachusetts Institute of Technology, Cambridge, Massachusetts 02139, United States}

\author{Alexandre C. Foucher}
\affiliation{Department of Materials Science and Engineering, Massachusetts Institute of Technology, Cambridge, Massachusetts 02139, USA}

\author{Michael Geiwitz}
\affiliation{Department of Physics, Boston College, Chestnut Hill, Massachusetts 02467, USA}

\author{Kenneth S. Burch}
\affiliation{Department of Physics, Boston College, Chestnut Hill, Massachusetts 02467, USA}

\author{Lukas Dekanovsky}
\affiliation{Department of Inorganic Chemistry, University of Chemistry and Technology Prague, Technická 5, Prague 6, 166 28, Czech Republic}

\author{Zdenek Sofer}
\affiliation{Department of Inorganic Chemistry, University of Chemistry and Technology Prague, Technická 5, Prague 6, 166 28, Czech Republic}

\author{Yu Liu}\altaffiliation{Present address: Los Alamos National Laboratory, Los Alamos,
NM 87545, USA.}
\affiliation{Condensed Matter Physics and Materials Science Division, Brookhaven National Laboratory, Upton, NY 11973, USA}

\author{Cedomir Petrovic}
\affiliation{Condensed Matter Physics and Materials Science Division, Brookhaven National Laboratory, Upton, NY 11973, USA}

\author{Frances M. Ross}
\affiliation{Department of Materials Science and Engineering, Massachusetts Institute of Technology, Cambridge, Massachusetts 02139, USA}

\author{Yimei Zhu}
\affiliation{Condensed Matter Physics and Materials Science Division, Brookhaven National Laboratory, Upton, NY 11973, USA}

\author{Prineha Narang}
\affiliation{Division of Physical Sciences, College of Letters and Science, University of California, Los Angeles, California 90095, USA}
\affiliation{Electrical and Computer Engineering Department, University of California, Los Angeles, California, 90095, USA}

\date{\today}

\begin{abstract}

Van der Waals (vdW) magnetic materials such as Cr$_2$Ge$_2$Te$_6$ (CGT) show promise for novel memory and logic applications. This is due to their broadly tunable magnetic properties and the presence of topological magnetic features such as skyrmionic bubbles. A systematic study of thickness and oxidation effects on magnetic domain structures is important for designing devices and vdW heterostructures for practical applications. Here, we investigate thickness effects on magnetic properties, magnetic domains, and bubbles in oxidation-controlled CGT crystals. We find that CGT exposed to ambient conditions for 5~days forms an oxide layer approximately 5~nm thick. This oxidation leads to a significant increase in the oxidation state of the Cr ions, indicating a change in local magnetic properties. This is supported by real space magnetic texture imaging through Lorentz transmission electron microscopy. By comparing the thickness dependent saturation field of oxidized and pristine crystals, we find that oxidation leads to a non-magnetic surface layer which is thicker than the oxide layer alone. We also find that the stripe domain width and skyrmionic bubble size are strongly affected by the crystal thickness in pristine crystals. These findings underscore the impact of thickness and surface oxidation on the properties of CGT such as saturation field and domain/skyrmionic bubble size and suggest a pathway for manipulating magnetic properties through a controlled oxidation process.

\end{abstract}
\maketitle

\section{Introduction}
The discovery of long-range ferromagnetic ordering in van der Waals (vdW) materials has opened new avenues for investigating fundamental magnetism and applications within novel memory, computing, and quantum computing \cite{gibertini2019magnetic, burch2018magnetism, mak2019probing, parkin2008magnetic, fert2017magnetic}. The tunability of vdW magnetic materials by thickness, i.e.\ number of atomic layers, as well as by electric fields, doping, and strain, and the presence of topological spin textures such as domain walls and skyrmions, make these materials promising for such applications \cite{gibertini2019magnetic, burch2018magnetism, mak2019probing}. However, an understanding of the evolution of spin textures as a function of flake thickness is highly desired.

\begin{figure*}
\scalebox{\figurescale}{\includegraphics[width=1\linewidth]{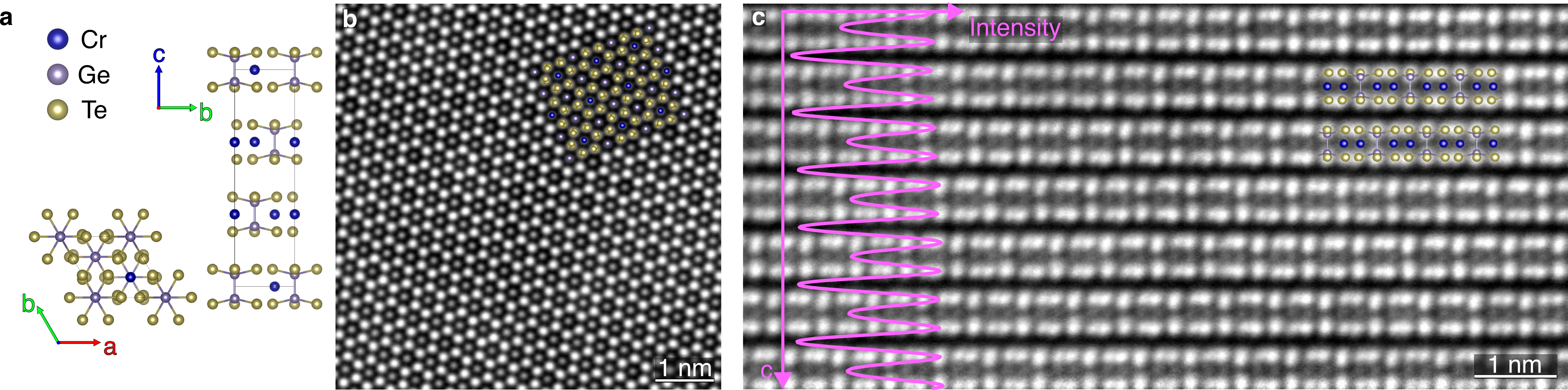}}
	\caption{\label{fig:sample}
        \textbf{Structure of CGT}
        \textbf{(a)} Atomic model of CGT. 
        \textbf{(b, c)} HAADF-STEM images of the CGT crystal in top view (b) and cross-sectional view (c). An atomic model of CGT is overlaid on both images. The grey-scale intensity of the image averaged horizontally is plotted on top of (c). 
		}
\end{figure*}

Cr$_2$Ge$_2$Te$_6$ (CGT) is one such material, a vdW Heisenberg semiconducting ferromagnet with a bulk Curie temperature of about 61~K  and strong perpendicular magnetic anisotropy \cite{gong2017discovery, zhang2016magnetic, carteaux1995crystallographic, vsivskins2022nanomechanical, xing2017electric}. The properties of CGT can be modified via doping
\cite{verzhbitskiy2020controlling}, electric field \cite{zhuo2021manipulating, xing2017electric, chen2018role}, strain \cite{vsivskins2022nanomechanical}, and pressure \cite{sun2018effects, lin2018pressure}. The latter two result from its strong magneto-elastic coupling \cite{tian2016magneto}. Because CGT is centrosymmetric it lacks a Dzyaloshinskii-Moriya interaction (DMI) to stabilize
Bloch or Néel-type skyrmions observed in noncentrosymmetric chiral magnets, such as B20 metals \cite{fert2017magnetic, nagaosa2013topological}. Instead, it hosts topological skyrmionic Bloch-type bubbles that are stabilized through a competition between the dipolar energy and magnetocrystalline anisotropy \cite{han2019topological, mccray2023direct}, making it a potential candidate for novel quantum computing~methods \cite{psaroudaki2021skyrmion}. 

The tunability of CGT and other vdW magnets stems from their large surface to volume ratio making the interfaces property-determining \cite{mak2019probing}. Proximitized spin-orbit coupling \cite{wu2020neel} and twisted homo- or hetero-bilayers \cite{tong2018skyrmions, hejazi2020noncollinear} 
are two interface effects that can tune magnetic properties of vdW magnets. A different approach may be to consider chemical effects, such as oxidation, where controlled oxidation can be used to create a chemically well-defined and non-reactive interface. CGT flakes are known to degrade in ambient air on the time scale of one hour \cite{gong2017discovery, tian2016magneto}. Consequently, to leverage this promising material in future applications, it is crucial to understand the effects of oxidation on its crystal and magnetic structure and assess possibilities for magnetic property control. 

Here, we prepare both pristine CGT and oxidized CGT (O-CGT) samples and determine how crystal thickness and oxidation affect magnetic properties and magnetic domain structures. Exposure to ambient conditions leads to the formation of a $\sim$5 nm thick oxide layer. By direct imaging of magnetic domains through Lorentz transmission electron microscopy (LTEM), we find a consistently lower saturation field for O-CGT compared to CGT. These measurements suggest that the effective magnetic thickness of O-CGT crystals decreases compared to CGT crystals and that the non-magnetic layer is thicker than the oxide layer alone. Compositional analysis using electron energy loss spectroscopy (EELS) and energy dispersive x-ray (EDX) analysis shows that the oxide has a lower Te content compared to bulk CGT, and that the oxidation state of Cr increases from 3+ in the bulk to a value between 3$^+$--\,5$^+$ in the oxide. Finally, having shown that oxidation leads to a decrease in the effective magnetic thickness, we consider how magnetic textures, which are useful for future applications, are affected by crystal thickness. We show that skyrmionic bubble sizes and stripe domain widths are tuned by crystal thickness. Our results thus show that deliberate chemical modifications of the crystal surface may be another avenue towards tuning magnetic properties of vdW magnets.

\section{Results and Discussion}

\subsection{Thickness dependent saturation field of CGT and O-CGT}

\begin{figure}
	\scalebox{\figurescale}{\includegraphics[width=1\linewidth]{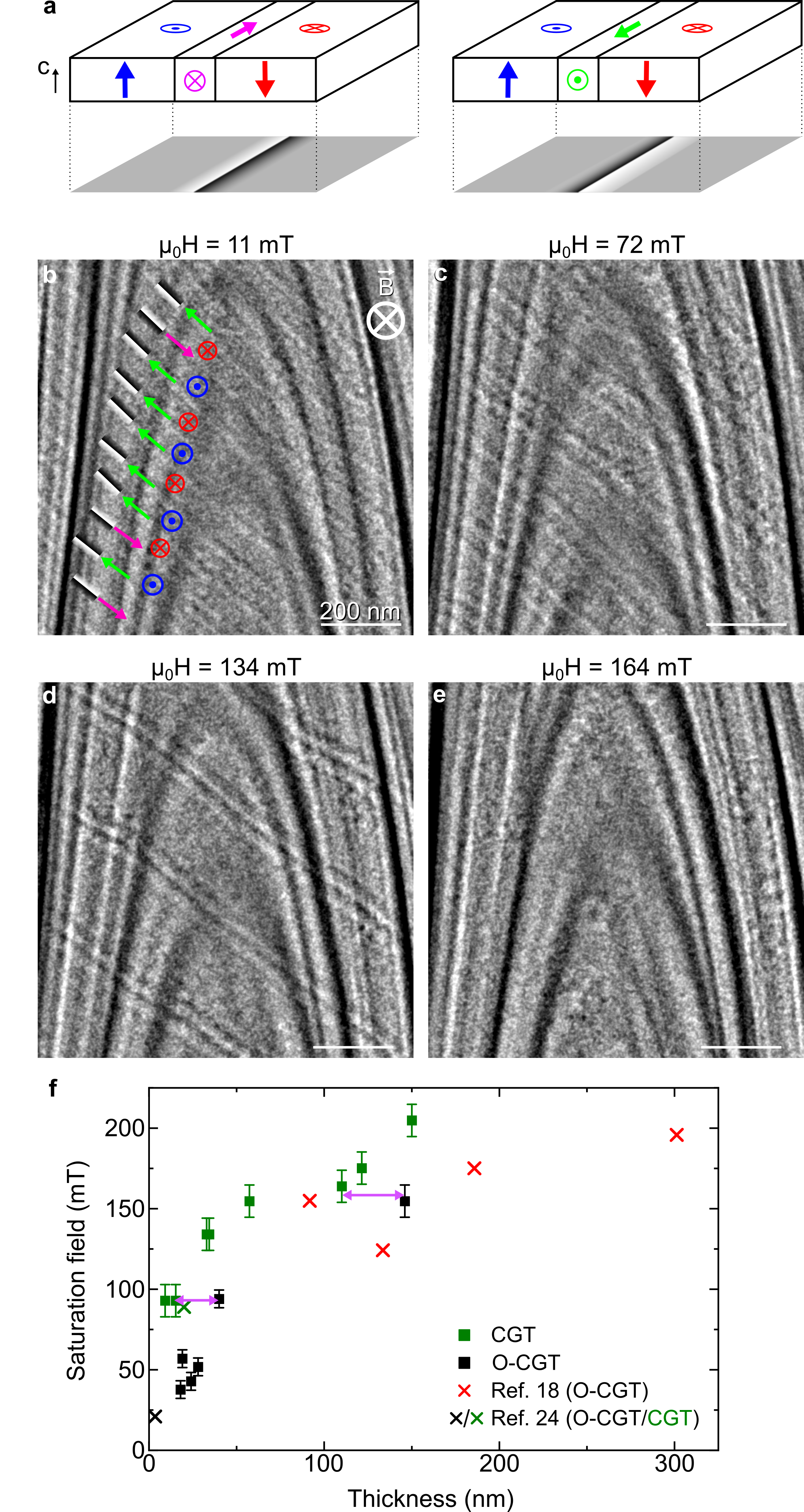}}
	\caption{\label{fig:crit}
		\textbf{Magnetic structure and saturation field.} 
        \textbf{(a)} Schematics of contrast formation in defocused LTEM imaging. Contrast is only formed by in-plane magnetization which leads to a phase shift in the electron beam, and either a dark-bright or bright-dark domain wall in the image plane, depending on the direction of the magnetization.
        \textbf{(b-e)} LTEM images after RFC at different magnetic fields for 110~nm thick CGT (measured by AFM). The saturation field is reached at 164~mT when there is no more magnetic contrast in the image. The stationary, and, therefore, non-magnetic, arch-shaped contrast in the images are bend contours. Overlays in (b) indicate the dark-bright or bright-dark contrast of the domain wall and the magnetization direction of the Bloch-type wall. 
        \textbf{(f)} Plot of the saturation field as a function of thickness and comparison with literature for oxidized and un-oxidized CGT. Red crosses are data extracted from Ref. \cite{han2019topological} (O-CGT), and black (O-CGT) and green (CGT) crosses are extracted from Ref. \cite{wang2018electric}. The pink arrows highlight the thickness difference between CGT and O-CGT crystals with similar saturation fields and have a width of 25~nm and 35~nm for lower and higher saturation fields, respectively.
  		}
\end{figure}

Crystal growth and sample fabrication are described in \textit{Methods}. CGT crystallizes in the trigonal R$\overline{3}$ space group (Fig.~1(a)) and the nominal oxidation states in bulk CGT are Cr$^{3+}$, Ge$^{3+}$, and Te$^{2-}$. Important for subsequent LTEM imaging, we confirm that the resulting samples are single crystalline and free from observable defects and interlayer dislocations, and do not degrade significantly during TEM observation as determined from high-angle annular dark-field scanning transmission electron microscopy (HAADF-STEM) images along the c-and a-axes (Fig.~1(b, c)), as well as TEM diffraction patterns (Fig.~S1). Figure~1(c) shows the vdW layered structure with ABC stacking, consistent with previous studies \cite{carteaux1995crystallographic, han2019topological, mccray2023direct}. Overlaid is an atomic model of CGT and the grey-scale intensity profile of the image averaged horizontally, from which we measure a layer thickness of 0.68~nm. Since the intensity in HAADF-STEM images is proportional to $Z^{1.7}$ \cite{hartel1996conditions}, $Z$ being the atomic mass, Te atoms appear the brightest.

In order to control the chemical state of the surface we use one of three approaches: (1) For O-CGT we prepare samples in ambient conditions. (2) For CGT we either (a) prepare samples in an argon-filled glovebox to avoid oxygen and moisture exposure \cite{tian2016magneto, gray2020cleanroom}, or (b) exfoliate CGT in ambient conditions but swiftly -- within minutes -- encapsulate between few-layer graphene. Details are provided in \textit{Methods}. Figure~S2 shows an illustration of sample details and optical contrast measurements and Fig.~S3 shows atomic force microscopy (AFM) measurements performed to measure crystal thicknesses. Briefly, bulk CGT is exfoliated, and suitable flakes are identified based on their optical contrast by first calibrating with measurements from AFM. As shown in Fig.~S2, the optical contrast can be used to estimate crystal thickness in the range 8-30 nm, while crystals with a thickness of 5-8 nm require AFM measurements because the contrast does not vary with thickness in this range. Pristine CGT samples prepared in the glovebox are stored within an air-sealed plastic bag, and therefore only exposed to air for some minutes before being loaded into the TEM. To form an oxide, we exfoliate and transfer flakes to TEM grids, and expose the samples to ambient conditions.

We first determine the saturation magnetic field in CGT and O-CGT for varying crystal thicknesses by imaging the magnetic domain structure at varying applied magnetic fields. Imaging is done using defocused LTEM with a liquid helium cooled holder with a base temperature of 15~K \cite{han2019topological}. In the LTEM mode the magnetic field at the sample can be controlled by exciting the objective lens. The magnetic field is oriented along the z-direction, along the electron beam direction, and the minimum achievable field in our TEM is 12~mT. Therefore, we refer to a cooling process from above the Curie temperature down to the base temperature of the sample holder at this magnetic field as “residual field cooling (RFC)” to distinguish it from a true zero-field cooling process. 

Contrast formation in small defocus ($<$ 1 mm) LTEM imaging is outlined in Fig.~2(a). In mechanically exfoliated CGT, magnetic domains magnetized parallel to the crystal c-axis separated by Bloch-type domain walls have been confirmed by previous studies \cite{han2019topological, mccray2023direct}. The out-of-plane easy axis of CGT has likewise been confirmed by several groups \cite{gong2017discovery, zhang2016magnetic, carteaux1995crystallographic}. The in-plane magnetization of a Bloch-type domain wall leads to dark-bright or bright-dark lines, depending on the direction of the in-plane magnetization. Hence, by considering the contrast of the magnetic features we can determine the direction of the domain walls. 

Our experimental procedure to determine the saturation field is shown in Fig.~2(b-e). After RFC, stripe domains have formed, with bright-dark pairs of lines indicating the domain walls (Fig. 2(b)). The domains are magnetized along the crystal c-axis, parallel to the microscope z-direction, as indicated in the image. Also indicated are the direction of magnetization of the domain walls as determined by the domain wall contrast. At the residual field, the domains have approximately equal widths. Note that these images also show more slowly varying contrast such as the arch-shaped contrast. This arises from bending of the sample and is common in magnetic imaging of vdW magnets \cite{han2019topological, mccray2023direct}. This contrast is stationary when we increase the magnetic field, and therefore, of non-magnetic nature.

As the magnetic field is increased (Fig.~2(c, d)), the magnetic domains aligned with the applied magnetic field grow in width while the magnetic domains aligned in the direction opposite to the magnetic field decrease in width. Finally, we reach the saturation field when we no longer can discern any magnetic contrast in the image (Fig.~2(e)).

\begin{figure*}
	\scalebox{\figurescale}{\includegraphics[width=0.85\linewidth]{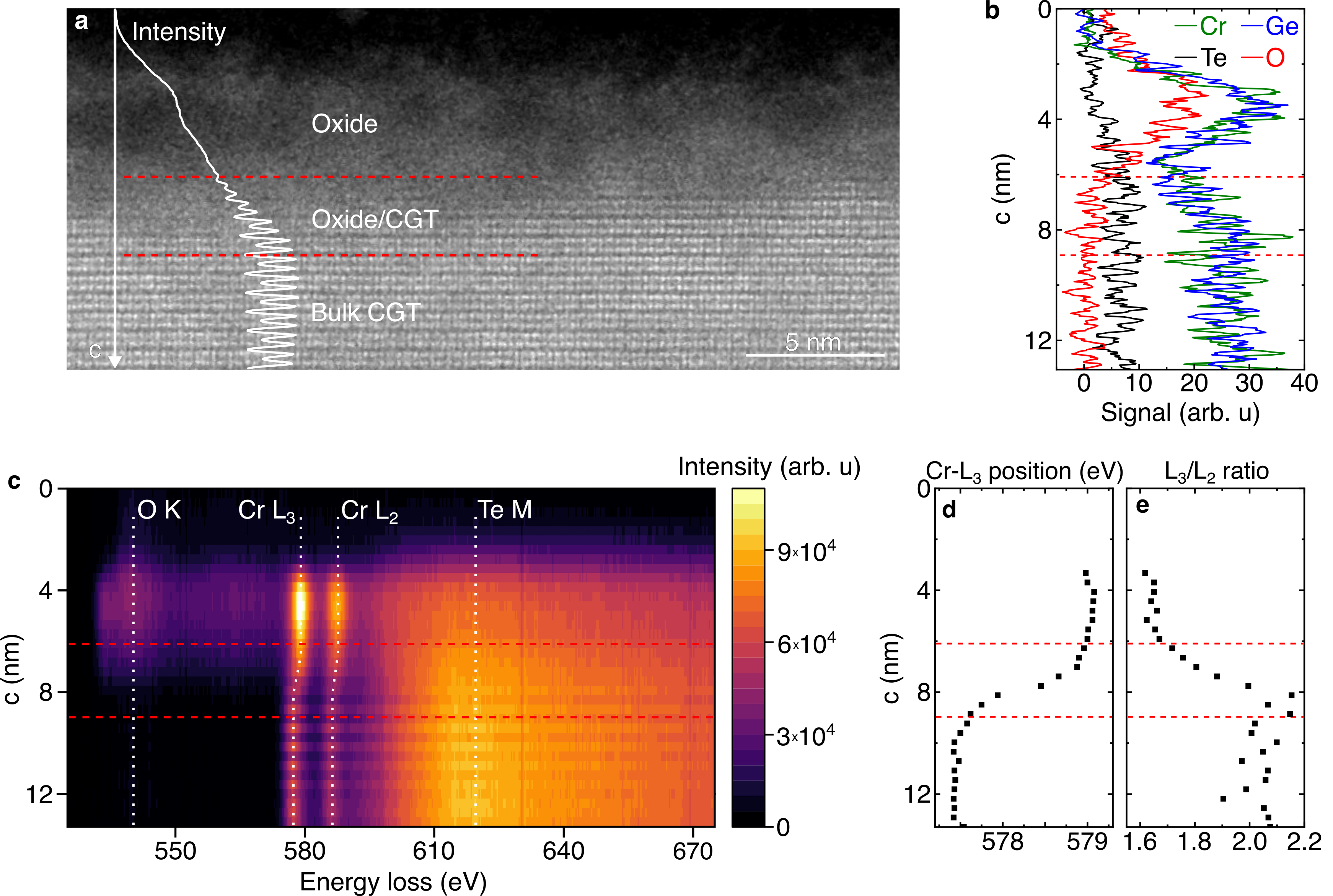}}
	\caption{\label{fig:EELS}
		\textbf{EELS and EDX analysis.} 
        \textbf{(a)} Cross sectional HAADF-STEM image of the surface of a CGT crystal that has been exposed to air for $\sim$30~days. The HAADF intensity is plotted on top and is averaged across the entire width of the image.       
        \textbf{(b)} EDX signal of Cr, Ge, Te, and O as a function of distance across the oxide/CGT interface, The EDX signal has been integrated horizontally for the width of the EDX maps shown in Fig.~S7 and smoothed using the Savitzky-Golay method with a 10-point window. The decrease in the Cr and Ge signal in the mixed oxide/CGT region may be due to variations in sample thickness, i.e., along the crystal b-axis. The lower signal at the mixed oxide/CGT region is seen in the EDX maps (Fig.~S7) and the EELS high-loss intensity (Fig.~S8(a)) both indicating a thinner sample in this region.     
        \textbf{(c)} EELS heatmap along the white line indicated in (c). Dotted white lines indicate the positions of the O K, Cr L$_2$/L$_3$ and Te~M peaks. The EELS signal has been integrated horizontally for the EELS maps shown in Fig.~S8. 
        \textbf{(d, e)} The position of the Cr-L$_3$ peak and the L$_3$/L$_2$ peak intensity ratio, respectively, through the oxide layer. The dotted horizontal red lines in (a-e) are at the same position in the crystal c-axis, at the interfaces between the oxide and mixed oxide/CGT region, and between the mixed oxide/CGT region and the bulk CGT. 
  		}
\end{figure*}

Figure~2(f) shows the saturation field plotted against crystal thickness for CGT and O-CGT. We find that the saturation field decreases with decreasing crystal thickness, and that CGT displays larger saturation fields compared to O-CGT at the same crystal thickness. This is consistent with prior studies of saturation field in CGT from which we were able to infer whether the CGT was oxidized or not \cite{han2019topological, wang2018electric}, also plotted in Fig.~2(f). We do not observe any qualitative difference in the morphology or appearance of the stripe domains between CGT and O-CGT other than the difference in the saturation field. By comparing data with similar saturation fields (see pink arrows in Fig.~2(f)), we infer that the effective magnetic thickness of O-CGT flakes is reduced. We estimate that O-CGT samples with a thickness around 50 nm have a combined non-magnetic layer of at least 25~nm, while O-CGT samples with a thickness of about 150~nm have a slightly thicker non-magnetic layer (at least 35~nm). We note that even though magnetic contrast decreases with thickness, it can still be discerned in O-CGT flakes down to 18~nm in thickness. We can, therefore, estimate a lower bound of $\sim$18~nm for the non-magnetic layer. Figure~S4 shows the magnetic structure of a 24~nm thick sample at varying magnetic fields.

\subsection{Oxide structure and composition}

To further understand the effect of oxidation on the surface structure and composition of CGT crystals we combine information from several measurements. We first quantify the crystallinity using HAADF-STEM cross-sectional imaging. Figure~3(a) shows the surface of a 100 nm thick CGT crystal viewed along the crystal b-axis. This flake was exfoliated onto a SiO$_2$/Si substrate and exposed to ambient conditions for $\sim$30~days before preparing the cross-sectional sample. At the top surface, we identify an amorphous oxide layer with a thickness of about 5~nm. The interface between the oxide and bulk CGT is not atomically sharp. On top of Fig.~3(a) we plot the image intensity averaged horizontally. From the intensity profile, we estimate a transition region that is 2-3~nm thick in which some crystallinity coexists with the amorphous oxide. The lower part of the image shows bulk CGT.

\begin{figure}
	\scalebox{\figurescale}{\includegraphics[width=1\linewidth]{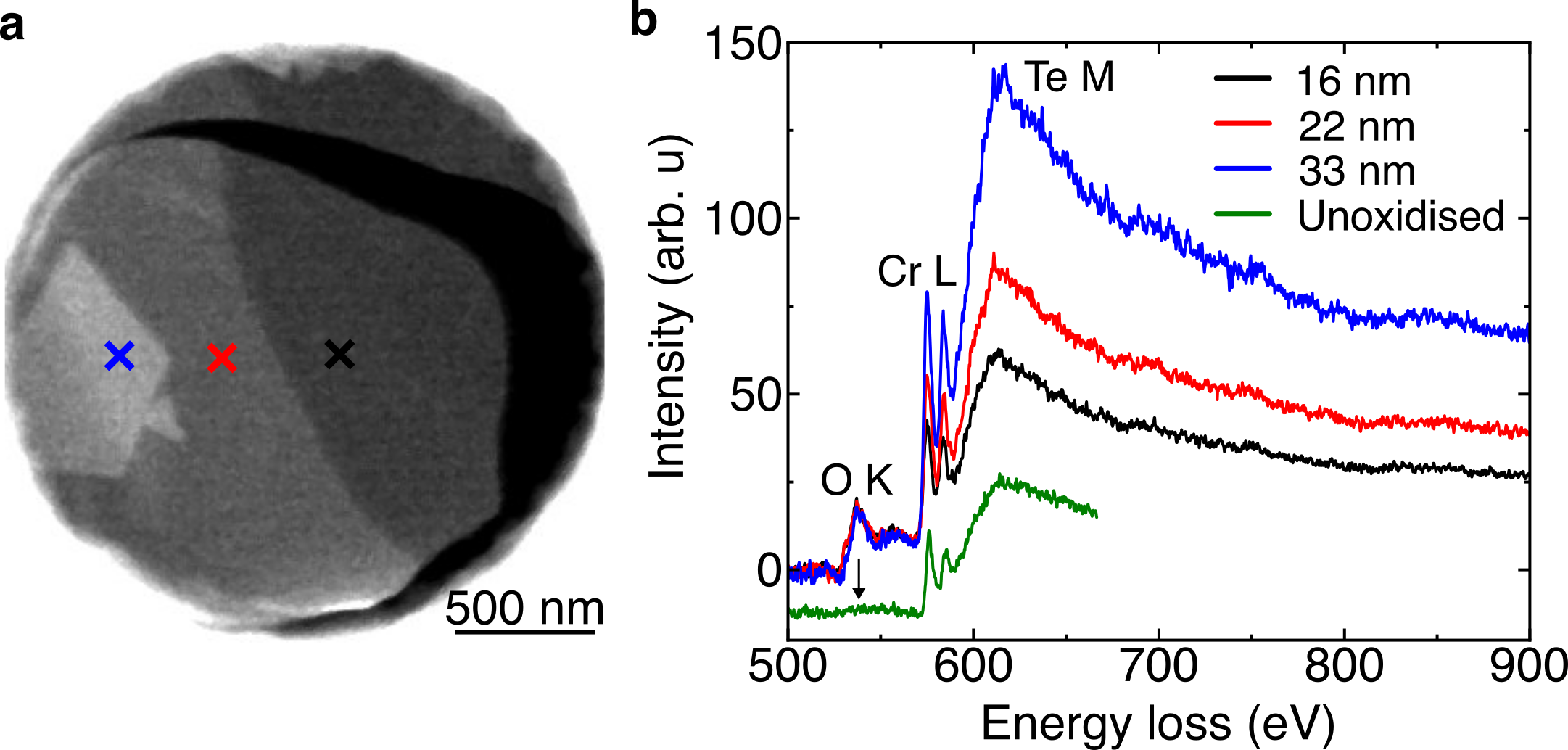}}
	\caption{\label{fig:domain}
		\textbf{Domain width analysis.} 
        \textbf{(a)} Plan-view STEM-EELS survey image of a CGT crystal with three different thicknesses. These sample thicknesses were determined using the low-loss EELS spectra with a mean free path for inelastic scattering of 76.4 nm – calibrated by AFM measurements. 
        \textbf{(b)} EELS spectra from the sample shown in (a) and a spectrum from a sample prepared in the glovebox showing the absence of the O K peak. The expected position of the O K peak is indicated by an arrow. The un-oxidized spectrum has been offset vertically for clarity. The black, red, and blue data are obtained from the areas denoted by crosses of the same colors in (a). The spectra have been deconvoluted with the low-loss EEL spectra. 
  		}
\end{figure}

Images of the bottom surface of the same sample (Fig.~S5) show an oxide thickness of 2-2.5~nm. This oxide is thinner because it experiences less oxygen exposure: it is exposed to atmosphere only for the few seconds it takes to cleave the crystal using scotch tape and place it on the SiO$_2$/Si substrate, although it is also exposed to oxygen that is trapped at the interface between CGT and SiO$_2$ after exfoliation. The two oxide thickness measurements indicate that oxidation of CGT initially proceeds quickly but then more slowly on further air exposure as it serves as a self-passivation layer. Furthermore, we have measured the oxide thickness of a CGT crystal that oxidized for 5 days (see Fig.~S6). This crystal had regions of different thicknesses of 35, 50, and 65~nm. Regardless of crystal thickness we measure a surface oxide thickness of about 5~nm. Thus, it can be concluded that the oxide thickness is self-limited, reaching its final thickness in less than 5~days, and independent of the CGT crystal thickness in the range 35-100~nm. 

We next investigate the elemental composition of the oxide and interface structure with EDX analysis. Figure~3(b) shows the EDX signal across the interface, acquired from a region similar to that shown in Fig.~3(a). Real space intensity maps of the Cr, Ge, Te, and O content obtained by EDX and EELS are given in Figs.~S7 and S8, respectively. In the oxide, the Cr, Ge, and O intensities track each other, suggesting that oxidation preserves the Cr/Ge composition and that the composition of Cr, Ge, and O is uniform. Furthermore, we find a decreased Te composition in the oxide compared to the bulk CGT. 

We then use electron energy loss spectroscopy (EELS) to investigate the Cr oxidation states across the interface. Figure~3(c) shows an EEL spectral map of the CGT surface shown in Fig.~3(a). We indicate the O K edge, the Cr L$_2$ and L$_3$ edge, and the Te M edge with dotted lines. The red horizontal dotted lines are in the same position as in Fig.~3(a,b) and indicate the transition between the oxide, mixed oxide/CGT, and bulk CGT. Consistent with our EDX data, we only detect an O signal at the surface of the CGT, spanning a region about 5~nm thick. We observe a 1.6-1.7 eV decrease of the Cr L$_2$ and L$_3$ peak positions. 
The L$_3$ peak position is plotted against c-position in Fig.~3(d) while the L$_3$/L$_2$ peak intensity ratio, also known as the white line intensity ratio, is plotted in Fig.~3(e). The background subtraction method for the L$_3$/L$_2$ ratio determination is based on a two-step function, as outlined in Fig.~S9 \cite{daulton2006determination}.

We find a Cr-L$_3$/L$_2$ ratio of about 2.0 in bulk CGT with nominal oxidation state of 3$^+$. This is slightly larger than typical values in literature \cite{daulton2006determination}. However, we note that quantitative comparison with published data is challenging due a large sensitivity to the method of background subtraction \cite{tan2012oxidation}. In our case, the presence of a Te edge at $\sim$620~eV affects the background after the Cr-L$_2$ edge. Focusing on relative shift in peak position and changes in L$_3$/L$_2$ ratio, we note that these take place in the region identified in Fig. 3(a) as a mixed oxide/CGT layer. Such a shift in peak position, decrease in L$_3$/L$_2$ ratio, and the presence of oxygen indicates an increase in oxidation state of the Cr \cite{daulton2006determination}. The maximum possible oxidation state of Cr is 6$^+$. However, based on the relative changes of these two parameters compared to literature values, we determine an upper bound of 5$^+$ for the Cr oxidation state \cite{daulton2006determination}. 

\begin{figure*}
	\scalebox{\figurescale}{\includegraphics[width=0.95\linewidth]{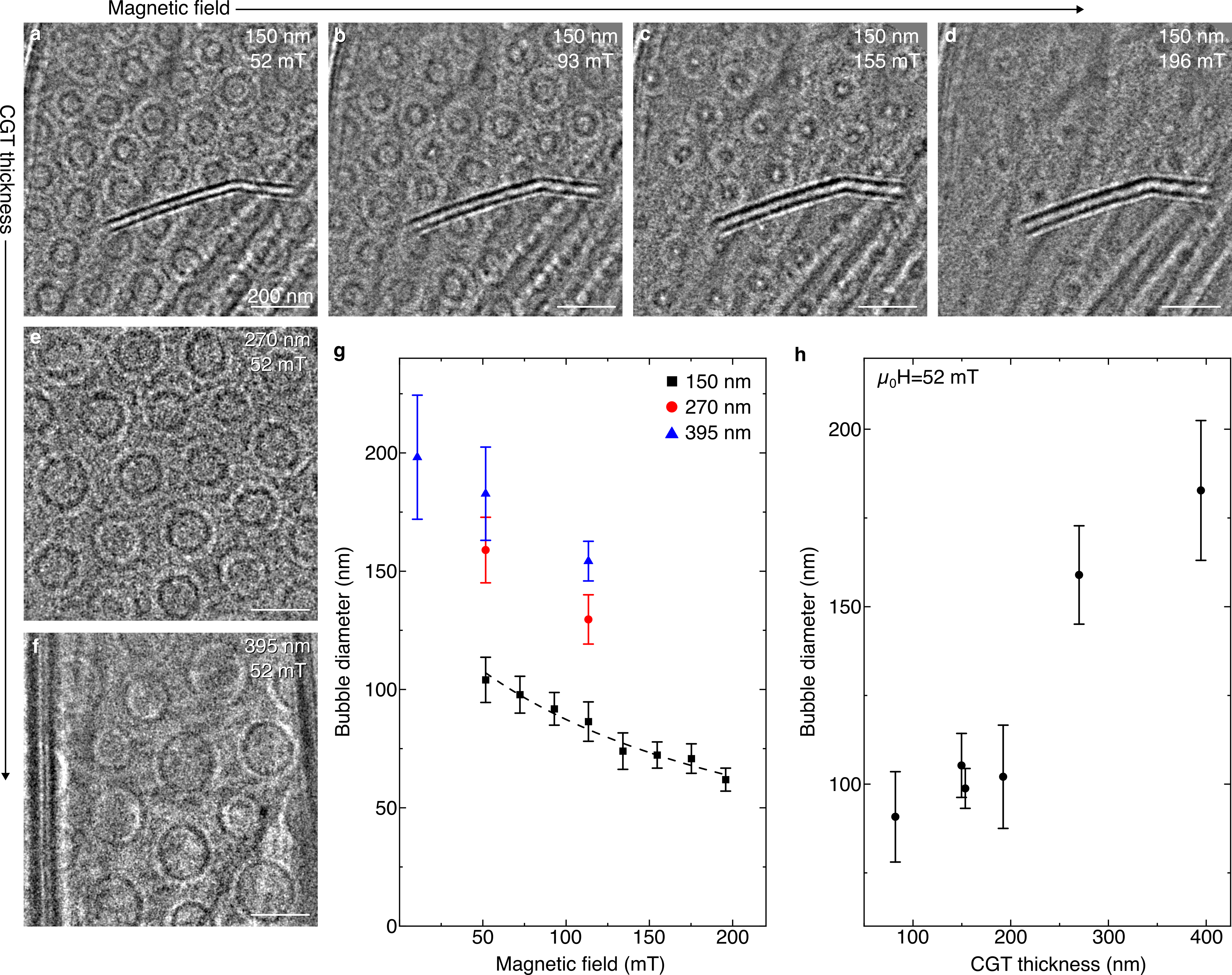}}
	\caption{\label{fig:bubble}
		\textbf{Control of skyrmionic bubble size.} 
        \textbf{(a-d)} LTEM images of skyrmionic bubbles in 150~nm thick CGT with varying magnetic field field from 52~mT to 196~mT. The image shown in (d) is the last one obtained before the saturation field of 204~mT. 
        \textbf{(a, e, f)} LTEM images at a constant magnetic field of 52~mT with increasing CGT thickness as indicated on the images. 
        \textbf{(g)} Bubble diameter plotted against magnetic field for three samples. The dashed black line is a fit to $A/(\mu_{0}H+B)$ where $A$ and $B$ are constants. 
        \textbf{(h)} Bubble diameter plotted against CGT thickness. In (g) and (h), the error bars are the standard deviations of the measurements of skyrmion diameter. All scale bars are 200~nm. The thickness of all CGT crystals has been measured by AFM. 
        }
\end{figure*}

Figure 4(a) shows an annular dark-field STEM image in plan-view, i.e., along the surface normal of the crystal, from an O-CGT sample with varying thickness. EEL spectra acquired from the regions indicated by the black (16~nm thick CGT), red (22~nm), and blue (33~nm) crosses are shown in Fig.~4(b). These spectra have been normalized by the oxygen K edge intensity. Since the EELS signal is proportional to the thickness of the sample for thin samples, these spectra suggest a constant oxide thickness independent of CGT thickness. This is supported by the additional measurements of oxide thickness from cross-sectional samples of varying thickness (Fig.~S6). Using the spectra in Fig.~4(a) we have quantified the relative compositions of O, Cr, and Te (see Fig.~S10 for fitting and results). As expected for a constant oxide thickness, the relative O content decreases with increasing crystal thickness. Furthermore, the relative Te content increases with increasing crystal thickness consistent with a Te deficiency within the oxide (Fig.~S10), consistent with our EDX measurements in Fig.~3(b). We note that sample thicknesses in Fig. 4 are all smaller than the mean free path for inelastic scattering and thus plural scattering effects on the composition measurement are likely not significant. Also shown in Fig.~4(b) is an EEL spectrum from a CGT sample prepared in a glovebox. This spectrum does not display an oxygen edge. Hence, our sample preparation protocol can achieve CGT samples with close to pristine surfaces due to the minimized air exposure. Samples prepared by exfoliating in ambient conditions but then rapidly encapsulating in graphene likewise do not display an oxygen peak (Fig.~S11). 

Overall, our results in Fig. 3-4 indicate that the non-magnetic surface layer is thicker than the amorphous oxide layer. We infer this because Fig. 2(f) suggests a non-magnetic surface layer that is at least 25 nm thick (top and bottom surfaces combined), while the high-resolution cross-sectional imaging in Fig. 3(a) and Fig.~S6 shows that the oxide thickness is about 5~nm. This suggests that both the oxide and the mixed oxide/CGT region are non-magnetic. This may be due to the change in oxidation state of the Cr which also changes the atomic spin moment and possibly the local magnetic interaction in the mixed oxide/CGT region and the oxide. Cr in bulk CGT has an oxidation state of  3$^+$ and an [Ar]3d$^3$ electronic structure with spin S=3/2, while a 4$^+$ or 5$^+$ oxidation state would correspond to an electronic structure of [Ar]3d$^2$ with S=1 or [Ar]3d$^1$ electronic structure with S=1/2, respectively.

Regarding our compositional findings as determined by EDX (Fig. 3(b)), the loss of Te in the oxide layer may contribute to the loss of magnetism observed in the oxide. This is because the superexchange interaction responsible for ferromagnetism in CGT is mediated between nearest neighbor Cr ions through the Te anions. The loss of Te in the oxide, furthermore, is similar to other vdW transition metal chalcogenide (TMD) compounds where the chalcogen is seen to be removed in the oxidation process, such as WSe$_2$ which forms WO$_3$ \cite{li2013mechanical} and MoS$_2$ which forms MoO$_3$ \cite{windom2011raman}. In such materials, oxide layer thickness and composition can be controlled using a thermal process or an oxygen plasma \cite{reidy2023atomic}, suggesting that for CGT it may also be possible to control the oxidation process.

\subsection{Thickness-dependent topological magnetic textures}

Having shown how oxidation changes the effective magnetic thickness of CGT crystals we next consider how magnetic textures change with thickness. Domain walls \cite{parkin2008magnetic} and skyrmions \cite{fert2017magnetic} hold potential for applications in emerging spintronic memory and logic devices. 

In Fig. 5 we explore the control of skyrmionic bubble size through CGT crystal thickness and magnetic field. LTEM images of CGT obtained after field cooling at a magnetic field of 52~mT results in skyrmionic bubble lattices, as previously reported for CGT \cite{han2019topological, mccray2023direct}. Increasing the magnetic field decreases the bubble size until the saturation field is reached. Figure~5(a-d) show images from a sample with a thickness of 150 nm. At 52~mT (Fig.~5(a)) the average bubble diameter is 104~nm: it decreases to 62 nm at a magnetic field of 196~mT (Fig.~5(d)), which is the last image obtained before the saturation field is reached. Figure~5(g) plots the bubble diameter against the applied magnetic field for three different samples. The size is inversely proportional to the applied magnetic field, as shown by the dashed black line in Fig.~5(g). This is as expected for skyrmions \cite{wilson2014chiral, romming2015field, yang2022magnetic}. 

Materials where magnetic bubbles are stabilized by a competition between magnetic dipolar interactions and magnetocrystalline anisotropy display thickness-dependent bubble sizes \cite{han2019topological, yu2014biskyrmion, ma2020tunable}. Conversely, Bloch-type skyrmions that are stabilized by a DMI have been reported to have a size that is independent of the sample thickness \cite{yu2011near, park2014observation}.
Figure~5(a, e, f) shows bubble lattices in several CGT crystals with different thicknesses, while Fig.~5(h) shows the bubble diameter plotted against CGT thickness, measured at a magnetic field of 52~mT. We find that the bubble diameter increases with CGT thickness (for example, diameter 90~nm at CGT thickness 82~nm; diameter 183~nm at CGT thickness 395~nm). This confirms that magnetic bubbles in CGT are stabilized by a competition between magnetic dipolar interactions and magnetocrystalline anisotropy, as expected in such a centro-symmetric material that lacks a DMI. 

Next, considering domain energetics, Fig.~6(a, b) show LTEM images obtained at 15~K after RFC for CGT crystals that are 15~nm and 150~nm thick, respectively. The 15~nm thick crystal gives less magnetic signal and we therefore use a larger defocus value to obtain measurable magnetic contrast. We used a lens in the image corrector system to control the defocus of LTEM imaging which is located far away from the sample and should therefore not affect the magnetic field at the sample plane. Figure~6(c) shows measurements of the domain width plotted against CGT thickness. We also plot data for Fe$_3$GeTe$_2$ (FGT) extracted from Ref.~\cite{li2018patterning}. As seen in this plot, FGT shows a similar behavior to CGT but displays larger domain width compared to CGT for the same crystal thickness. The two materials are compositionally analogous and both have been shown to oxidize in ambient conditions leading to a non-magnetic surface layer \cite{li2023visualizing, kim2019antiferromagnetic}.
This suggests that the findings here may be more widely applicable to other vdW magnets. 

\begin{figure*}
	\scalebox{\figurescale}{\includegraphics[width=1\linewidth]{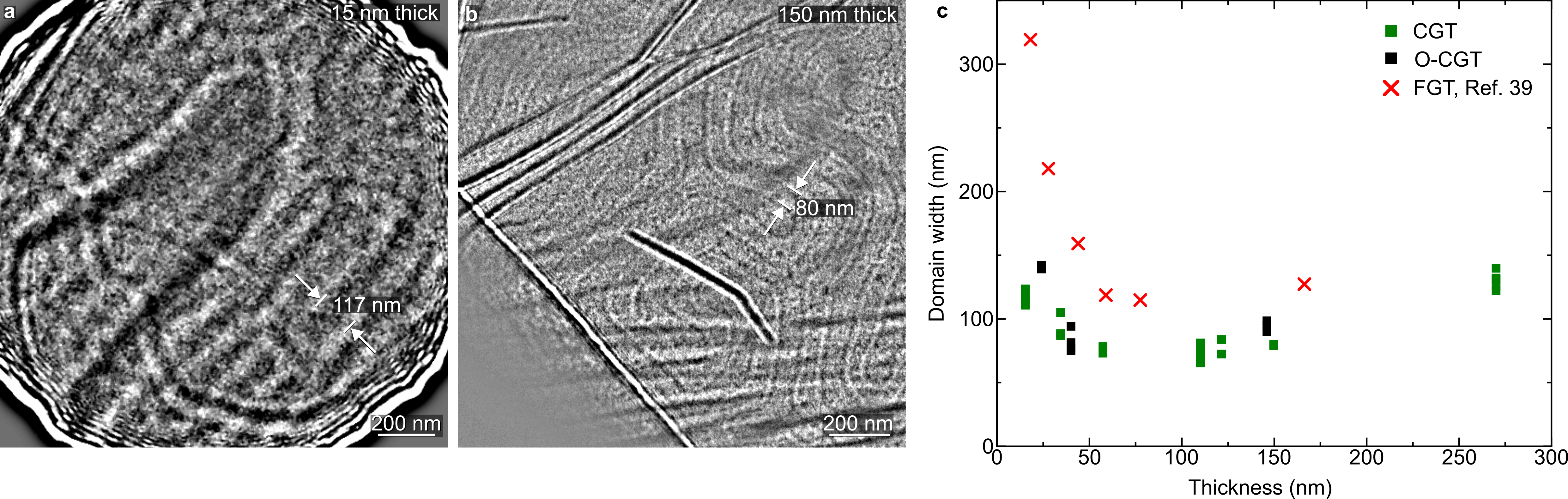}}
	\caption{\label{fig:domain}
		\textbf{Domain width analysis.} 
        \textbf{(a, b)} LTEM images of magnetic domain structure in (a) 15 nm thick and (b) 150 nm thick un-oxidized CGT (thickness measured by AFM). The SiN membrane is visible in the corners of (a). The indicated domain widths represent the average domain width measured in the image. 
        \textbf{(c)} Domain width as a function of CGT thickness. Red crosses are extracted from Ref.~\cite{li2018patterning} for Fe$_3$GeTe$_2$. Data was obtained at the smallest possible magnetic field of 11 mT.
  		}
\end{figure*}

Kittel developed a model for the energetics of stripe domains in thick films of ferromagnetic material with out-of-plane magnetic anisotropy \cite{kittel1946theory}. Kooy and Enz later extended this model to include thin membranes \cite{kooy1960philips}. The domain width depends on a competition between the domain wall energy, anisotropy energy, and the demagnetizing energy (stray field energy). For very thin membranes, the domain width increases exponentially with decreasing membrane thickness. As the thickness is increased, a minimum domain width, $L_{\mathrm{min}}$, is reached at a thickness of $4\sigma_{\mathrm{w}}/\mu_{0}M_{\mathrm{s}}^2$, where $\mu_{0}$ is the vacuum permeability, $\sigma_{\mathrm{w}}$ is the specific domain wall energy, and $M_{\mathrm{s}}$ is the saturation magnetization (Sec. 3.7.3 in Ref. \cite{hubert2008magnetic}). For CGT and FGT, $L_{\mathrm{min}}$ is reached at thickness of $t_{\mathrm{D}}$ = 60-120~nm and $t_{\mathrm{D}}$=80-160~nm, respectively. By using $M_{\mathrm{s,CGT}}=1.37\cdot10^5$ A/m and $M_{\mathrm{s,FGT}}=3.76\cdot10^5$~A/m from Ref.~\cite{wu2022van}, we can estimate the domain wall energy in CGT and FGT to be in the range (4-7)$\cdot 10^{-4}$~J/m$^2$ and (36-71)$\cdot 10^{-4}$~J/m$^2$, respectively. This implies that the domain wall energy is around one order of magnitude smaller in CGT compared to FGT. The latter value fits well with another estimate of 47$\cdot 10^{-4}$~J/m$^2$ for the domain wall energy of FGT \cite{leon2016magnetic}. We note that Refs.~\cite{tian2016magneto, mccray2023direct} found sizable magnetostrictive effects in CGT, and therefore a magnetostrictive term should possibly be included in the Kooy-Enz model for completeness. 

Based on the Kooy-Enz model, we would expect CGT to display behavior similar to FGT for thin layers with thickness $< t_{\mathrm{D}}$, where the domain width increases until a critical thickness is reached and finally the sample becomes magnetized in a single domain. Measurements of ultra-thin (6 layers thick) crystals with lateral dimensions $\sim$10 $\mu$m have previously been shown to have a single magnetic domain \cite{gong2017discovery}.
Additionally, theoretical studies have predicted that CGT hosts a thickness-dependent magnetic anisotropy which decreases for thin crystals \cite{fang2018large, song2019surface}. Thus, one would expect the domain wall width, $W_d$, to increase for thin crystals, since $W_{d} \sim \sqrt{A/K}$, where $A$ is the exchange constant and $K$ is the anisotropy constant \cite{hubert2008magnetic}. We measure the domain wall width for three relatively thick crystals, with thickness from 110 to 390~nm, to be in the range 6-14~nm (Fig.~S12), without any clear thickness dependence. This indicates that the uniaxial anisotropy remains robust for this range of thicknesses. For thin oxidized crystals (thickness = 24 nm) we do not observe a stripe domain structure that is as well-defined as those shown in Fig. 6 (Fig.~S4), which suggests a weaker magnetic anisotropy. Hence, it is important to take oxidation into account when predicting the properties of a CGT crystal. Finally, future studies should determine whether a minimum thickness exists for the presence of domain walls and skyrmionic bubbles in CGT. 

\section{Conclusions}
Our results show how oxidation yields a non-reactive and chemically well-defined surface, serving as an additional control knob for magnetic properties in CGT. This may be applicable to vdW magnets more generally, where control of the oxide growth may be achievable through thermal or plasma oxidation -- as shown for TMDs \cite{reidy2023atomic} and with layer-by-layer control \cite{zheng2018controlled}. 

We found that exposure to ambient conditions for 5~days results in a non-magnetic, self-limited surface oxide layer that is about 5~nm thick. Measurements of saturation field as a function of crystal thickness for CGT and O-CGT crystals suggest that oxidation reduces the effective magnetic thickness. The thinnest oxidized crystal in which we observed magnetic textures was 18 nm thick, giving an estimate of the minimum non-magnetic surface layer. From EELS analysis we determined an upper bound of 5$^+$ for the Cr oxidation state in the oxide which increased from 3$^+$ in the bulk, indicating a change to the local magnetic properties. 

We further observed a strong thickness dependence on the size of skyrmions which suggests that chemical surface control can in principle tune skyrmion sizes. Additionally, oxidation could be used for patterning, as recently demonstrated for TMDs \cite{kim2023ambipolar}. With increasing knowledge of the electrical and magnetic properties of the surface oxide layer, and further evaluation of the degree to which its growth can be controlled in a manner analogous to the growth of oxides such as MoO$_3$ and WO$_3$, we speculate that surface chemical control may become a useful tool for developing electrical or magnetic applications of 2D magnets.

\section{Methods}
\subsection{CGT crystal growth}
Bulk CGT used for the black data points in Fig.~2(f) and Fig.~6(c) was grown by the self-flux technique starting from a mixture of pure elements: Cr (99.95\%, Alfa Aesar) powder, Ge (99.999\%, AlfaAesar) pieces, and Te (99.9999\%, Alfa Aesar) pieces. A molar ratio of 1:2:6 for Cr:Ge:Te was used. The starting materials were sealed in an evacuated quartz tube, which was heated to 1100\,$\degree$C over 20~hours, held at 1100\,$\degree$C for 3~hours, and then slowly cooled to 700\,$\degree$C at a rate of 1\,$\degree$C/hour. 

Bulk CGT used for the remaining data was prepared by direct reaction from elements with an excess of Te. The high purity Te (99.9999\%), Ge (99.9999\%) and Cr (99.99\%) were mixed in a quartz ampoule with a stochiometric ratio of 2:2:10. The 50~g of reaction mixture was sealed under high vacuum in a quartz ampoule (25x150~mm, 3~mm wall thickness) and heated in a muffle furnace for 6~hours at 1000\,$\degree$C with heating rate of 2\,$\degree$C/min. The reaction mixture was horizontally placed in furnace and  mechanically shaken several times at 1000\,$\degree$C. The ampoule was cooled to 450\,$\degree$C at a rate of 6\,$\degree$C/hour and then left to cool freely to room temperature. The ampoule was opened inside an argon glovebox and CGT crystals with size exceeding 5x5 mm were mechanically separated from the middle of the molten ingot.

\subsection{Sample fabrication}
CGT crystals were obtained by exfoliating bulk CGT using scotch tape onto substrates of 90~nm thick SiO$_{2}$ on Si using 3M Magic Scotch tape. Oxidized crystals were exfoliated in ambient conditions while un-oxidized crystals were exfoliated in a glovebox. Some un-oxidized samples were prepared by exfolating the CGT in ambient conditions and immediately (within $\sim$10 minutes) assembling the CGT into a van der Waals (vdW) heterostructure with graphene encapsulation. EEL spectra of the vdW heterostructures do not display an O peak indicating the absence of any measurable oxidation (Fig.~S11). This is in line with previous reports of oxidation taking place on the timescale of an hour \cite{gong2017discovery}.

Crystals and vdW heterostructures were then transferred to location-tagged TEM grids purchased from Norcada, Inc., Canada. The transfer was performed using wedging transfer in which a film of cellulose acetate butyrate is used as polymer handle \cite{thomsen2022suspended, thomsen2019oxidation}. After the transfer we bake the TEM sample carriers to 80\,$\degree$C to improve adhesion. Finally, we dissolve the in acetone and rinse the TEM sample carriers in isopropanol.

Cross-sectional samples were prepared by focused ion beam milling using a Helios Nanolab 600 DualBeam instrument (FEI). A protective layer of Pt was first deposited with the electron beam (at least 2~$\mu$m thick) before the cut and the lift-out. The lamella was then attached to a half-grid and thinned down to obtain a thickness of 100~nm. The last step of the thinning process was performed with the ion beam operating at 5~V with a current of 0.15~pA to avoid damage. Both cross-sectional samples (one used for Fig. 3, S7, and S8 and one used for Fig.~S6) were stored in a vacuum desiccator and imaged in the STEM less than 24 hours after their preparation. 

\subsection{Electron microscopy}

A JEOL ARM 200CF instrument equipped with cold field emission gun and double-spherical aberration-correctors at Brookhaven National Laboratory was used for LTEM and the EELS data shown in Fig.~3(g). A double-tilt liquid helium cooling holder (HCTDT 3010, Gatan, Inc.) was used for low-temperature experiments. The magnetic field at the sample plane in the LTEM mode as a function of objective lens strength has been calibrated using a modified TEM holder with a Hall-probe \cite{lau2007straightforward}. 

HAADF-STEM imaging and EELS data (for Fig.~3(a, c-e) was acquired  on a probe-corrected Thermo Fisher Scientific Themis Z G3 60-200~kV S/TEM operated at 200~kV. EELS data was acquired with a Continuum EEL spectrometer and a dispersion of 0.3~eV/channel, a dwell time of 0.02~s, beam current of 50~pA, and a zero loss peak full-width half maximum of 1.9~eV. The data was denoised using principal component analysis through the implementation in Hyperspy \cite{hyperspy}. The EELS maps shown in Fig.~S8 from which the data in Fig. 3(c-e) is derived, are 36x144 pixels large, with a pixel size of 0.37 nm. Thus, in Fig. 4(c), the resolution on the c-axis is also 0.37 nm/pixel. The cross-sectional samples were stored in a vacuum desiccator and imaged the day after FIB preparation.

EDX data was aqcuired with a Hitachi HF5000 environmental TEM operated at 200~kV with an electron beam current of 100 pA and a dwell time of 10 µs. The acquired EDX maps (Fig.~S7) are 630x455 pixels large, with a pixel size of 0.31 Å. The line profiles shown in Fig. 3(b) also have a resolution on the c-axis of 0.31 Å and were extracted from the EDX maps using the “TruLine” option, also known as the “Filtered Least Squares”, in the AZtec EDS software from Oxford Instruments. EDX measurements were performed 5 days after FIB sample preparation.

\subsection{Data analysis}
From each LTEM image, a 40~pixel Gaussian blurred version of the image  was subtracted to remove long-range intensity variations, followed by the application of a median filter. Domain width was measured by selecting several neighboring domain walls and measuring their average distance. Magnetic bubble diameters are measuring by tracing out their perimeter, and calculating the corresponding diameter assuming a perfect circle. For both bubbles and domain walls, the position of the domain wall is determined as the transition between bright and dark contrast on the LTEM images. 

\subsection{Atomic force microscopy}
AFM was performed on an Asylym Research Zypher VRS AFM in tapping mode. 

\section{Acknowledgements}
This work is primarily supported through the Department of Energy BES QIS program on "Van der Waals Reprogrammable Quantum Simulator" under award number DE-SC0022277 for the work on long-range correlations, as well as partially supported by the Quantum Science Center (QSC), a National Quantum Information Research Center of the U.S. Department of Energy (DOE) on probing quantum matter. M.G. acknowledges support by the Air Force Office of Scientific Research under award number FA9550-20-1-0246. The work at the Brookhaven National Laboratory was supported by the U.S. Department of Energy (DOE), Basic Energy Sciences, Materials Science and Engineering Division under Contract No. DESC0012704. K.S.B. acknowledges the primary support of the DOE, Office of Science, Office of Basic Energy Sciences under award number DE-SC0018675. The authors acknowledge the use of the MIT.nano Characterization Facilities. The authors thank Caroline A. Ross for fruitful discussions. Figure 2(a) was created using Vesta \cite{momma2011vesta}. P.N. gratefully acknowledges support from the John Simon Guggenheim Memorial Foundation (Guggenheim Fellowship) as well as support from a NSF CAREER Award under Grant No. NSF-ECCS-1944085.

\bibliographystyle{achemso}
\bibliography{references}

\end{document}